\documentclass{IEEEtran}
\usepackage{times}  %
\usepackage{helvet}  %
\usepackage{courier}  %
\usepackage{url}  %
\usepackage{graphicx}  %
\usepackage{biblatex} 

\bibliography{farnan_wtmc}

\ifCLASSINFOpdf
\else
\fi

\begin{document}
\title{Analysing Censorship Circumvention with VPNs via DNS Cache Snooping}

\author{\IEEEauthorblockN{Oliver Farnan}
\IEEEauthorblockA{University of Oxford\\
oliver.farnan@cs.ox.ac.uk}
\and
\IEEEauthorblockN{Alexander Darer}
\IEEEauthorblockA{University of Oxford\\
alexander.darer@cs.ox.ac.uk}
\and
\IEEEauthorblockN{Joss Wright}
\IEEEauthorblockA{Oxford Internet Institute\\
joss.wright@oii.ox.ac.uk} }

\maketitle

\begin{abstract}

Anecdotal evidence suggests an increasing number of people are turning to VPN services for the properties of privacy, anonymity and free communication over the internet. Despite this, there is little research into what these services are actually being used for. We use DNS cache snooping to determine what domains people are accessing through VPNs. This technique is used to discover whether certain queries have been made against a particular DNS server. Some VPNs operate their own DNS servers, ensuring that any cached queries were made by users of the VPN. We explore 3 methods of DNS cache snooping and briefly discuss their strengths and limitations. Using the most reliable of the methods, we perform a DNS cache snooping scan against the DNS servers of several major VPN providers. With this we discover which domains are actually accessed through VPNs. We run this technique against popular domains, as well as those known to be censored in certain countries; China, Indonesia, Iran, and Turkey. Our work gives a glimpse into what users use VPNs for, and provides a technique for discovering the frequency with which domain records are accessed on a DNS server.

\end{abstract}

\IEEEpeerreviewmaketitle

\section{Introduction}

Increasing global government crackdowns on internet privacy, anonymity and free communication have found users looking for technologies which they believe can help restore these properties. One of these is the Virtual Private Network (VPN). While not originally designed for this purpose, nevertheless, many have turned to them, often over technologies specifically designed with such properties in mind e.g. Tor \cite{dingledine2004tor}, Psiphon \cite{psiphon}, and Signal \cite{signal}\cite{cohn2017formal}. VPNs allow users to trivially, quickly and cheaply make it appear as if they are connecting from a different IP address and location, giving users a sense of anonymity and allowing access to content that may not be accessible through their regular internet connection. They also encrypt traffic between the client and VPN  gateway, preventing or hindering network traffic analysis or surveillance. They are in widespread use in countries with strict internet censorship regimes \cite{nabi2013anatomy}\cite{aryan2013internet}, and they are being increasingly targeted to prevent this use. Most VPN providers advertise on the basis of evading censorship and surveillance \cite{bypassrussia}\cite{ipvanishbypass}\cite{vpnguru}, and there has been academic work either implicitly or explicitly assuming this connection \cite{papadopoulos2018exclusive}\cite{rainie2013anonymity}\cite{savchenko2015analytical}. Possibly the largest internet censorship regime on the planet is currently attempting to block, remove or regulate their use \cite{chinavpn}\cite{applechinavpn}.

Despite this, VPNs are not always designed and implemented with strong security, anonymity or privacy guarantees. Due to the ease of setting up a VPN, there are many organisations offering both free and paid VPN services. Many of these are fairly basic, and cannot offer the same level of security as technologies which are designed with these requirements in mind \cite{appelbaum2012vpwns}\cite{perta2015glance}. 

VPNs create an encrypted tunnel between the client and the VPN gateway. Once this tunnel is established, regular IP routing is used to make requests such as accessing web content. This requires access to Domain Name System (DNS) servers.  There is no standardised way for VPNs to handle DNS infrastructure, and so most VPNs handle these requests in their own way: some making no provision for DNS requests, others forwarding requests onto public servers such as 8.8.8.8 \cite{googlepublicdns}, and others still operating DNS servers themselves. Over time, users of VPN services seeking privacy and anonymity have become aware of the risks of deanonymisation through DNS requests. As a result, an increasing number of VPN services now include a DNS resolution service internal to the VPN. However these DNS servers can reveal information about the users and uses of these VPN services.

\section{DNS Cache Snooping}
DNS cache snooping is a technique that allows the discovery and analysis of DNS records that a particular server has cached. While it cannot (ordinarily) be used to breach the privacy of individual users, it can be used against services to discover what users are collectively using the service for. While a known technique, there is little academic work focusing on it  \cite{grangeia2004dns}. Using DNS cache snooping it is possible to perform an analysis on which domains are being queried against that server and how frequently they are being accessed.

Private VPN subscriptions typically come with an application or OpenVPN \cite{openvpn} configuration file which makes changes to the network interface of the client host. There has been some concern in the VPN community about `DNS leaks', where user anonymity is compromised by DNS requests sent to third party DNS servers \cite{nordvpnleak}\cite{torguardleak}\cite{bestleak}. In response, VPN services are increasingly offering their own internal DNS server to prevent requests leaving the network. Normally these servers are only available once connected to the VPN\footnote{During our analsysis we found one instance of a possibly misconfigured VPN DNS server which was accessible over the open internet.}.

For this work we focused on those VPN providers which offered their own DNS servers. Once we were connected to the VPN we made repeat DNS queries for domains we were interested in. There were three possible ways to carry out DNS cache snooping scan from this position.

\textbf{TTL with no recursive queries }--  The first method is to query a DNS server with the Recursion Desired flag set to 0. This flag is intended to prevent the server from performing a recursive lookup in the case that the entry is not cached locally. If no cached record is returned then we know that no user has made a recent query for this domain or that the DNS server does not return anything when RD is 0. This period is determined by the maximum DNS TTL setting which is configured on the authoritative DNS server for that domain. This value is set in seconds, and states the amount of time a caching DNS server should hold that record before letting it expire. Once it expires, a new request must be sent to get a fresh record. Unfortunately, DNS servers are not required to respect the Recursion Desired flag. Upon receiving these requests some servers perform a standard recursive lookup for that domain. While servers respecting this flag can be found on the open internet, for the purposes of this work we did not find any internal VPN DNS servers that did so.

\textbf{TTL with recursive queries }-- The second method is similar but does not rely on the Recursion Desired flag. Instead we wait for the cached record to expire, and then wait for an additional period no longer than the maximum observed TTL for that domain, before making our own query. By subtracting the remaining TTL from the query from the maximum TTL for that domain, we see how long it took for the record to be refreshed after expiring.
  
This method relies on knowing the maximum TTL for each record on that server. TTL values are set by the authoritative server for that domain, but we found in practice that some DNS servers do not respect this. In response, a precursor step to this approach is discovering the maximum TTL for a domain on a per server basis.

This approach is less ideal than the first as our queries pollute the DNS server's cache. Because of this there are periods of time where we are unable to observe queries for this domain: as the records are cached and counting down from our request. As a result more data is needed to get the same accuracy in results. However this approach is reliable and works on almost all DNS servers we tested against\footnote{The only server did not work on was the aforementioned server that was refreshing its cache before the TTL had expired}, and is the method that we ultimately used

\textbf{Time based }-- The third method of DNS cache snooping is timing based, depending on whether the query is cached on that server, or if the server has to make a recursive lookup for this domain. This technique is ideal for highly frequented domains or active DNS servers, and can identify granularities of frequency of less than one second. Unfortunately this approach is susceptible to network jitter and fluctuating delay variations caused by third party DNS services and CDNs. We found that although there is a large increase in response time when making a request for a domain that is not cached, such response time differences were common. The majority of requests we made that took an increased amount of time to respond to were unpredictable jitter, rather than recursively retrieving a record from another server. This is likely not an insurmountable problem, and can probably be overcome with enough data and noise filtering techniques.

\section{Methods}

We ran our experiment against three internal DNS servers belonging to popular subscription VPN services. To make requests to these services a client had to be connected to the VPN. Once connected they were able to make requests and perform DNS cache snooping to see which domains were being accessed. All of these experiments were run during April and May 2018. 

\textbf{Popular Domains} -- The first stage of our experiment was run against 1000 popular domains. This list was taken from the Majestic Million \cite{majestic}. We choose this top domains list as it is based on web backlinks, similarly to how we choose our most popular censored domains (described below) \cite{scheitle2018long}. 11 of these domains had expired or were not live \cite{bano2018scanning}, so these domains were removed and replaced with new domains from the list to keep the total at 1000. These inactivate domains were primarily service subdomains which had recently been disabled and were no longer in use.

\textbf{Censored domains} -- The second stage of the experiment we repeated against popular domains that we knew to be censored. We gathered these domains using the technique described by Darer et al. \cite{darer2018automated}\cite{darer2017filteredweb}. This list was also based on backlinks from web pages, similar to the Majestic Million, and so domains with more links to it from other sites were chosen over those domains that only had a few backlinks. Domains which were duplicates between the lists were removed, as were domains that had expired and consistently failed to resolve to an IP address.

\textbf{Language specific censored domains} -- For the third and final stage of the experiment we ran the process against language specific domains. One weakness of our method is that we are unable to tell who accessed a record for a domain: only that it was accessed. Many of the domains in these lists contained content that is relevant to a wide variety of nationalities, for example facebook.com or blogger.com. These platforms allow the publication  of a wide variety of content, and it is impossible for us to say whether someone is accessing this from say Indonesia (or, to access Indonesian related content), or from another country entirely.

With this in mind, and to find domains where the content was specific to the four countries that we were looking at, we decided to separate these domains by language. From here we classified domains based on 1) the language of the content of the landing page, and 2) the HTML ISO language code. Domains where these two pieces of information gave conflicting reports had their content checked manually. Domain content classification was performed with the Google Compact Language Detector v3 (CLD3) open library \cite{cld3}. After language classifying the web content of the pages we removed the domains where the content was primarily in a language that was different to the country it was censored from. For example, we found that shahrvand.com was censored in Iran, and that its primary language was Farsi, so we kept it in the language specific list. Blogger.com contained a large variety of languages and was not specific to any one country, and so was removed. Using this method we ensured we kept only those domains which contained content that was directly related to the country that had restricted access to it. As these domains were known censored in their home countries, and as the language of these pages is specific to those countries, using accessing these domains through the VPN were likely to be doing so for the purposes of censorship evasion.

Initially we tried to query the authoritative servers for each domain to discover their maximum TTL, but we found that some servers did not respect this. For these domains we had to manually discover the maximum TTLs being used for each server domain pair. We began this process by sending a single query for each domain. This query returned a TTL value stating when the record was due to expire. We then repeatedly polled the DNS server for that domain as the TTL value counted down. When this value reached 0 it would roll over and return the suspected maximum TTL. This process was rerun until the same maximum TTL had been seen for a given domain 5 times, to ensure we had the correct value. In practice the first observed suspected maximum TTL was the finally accepted value in $>$95\% of cases. Additionally, all observed maximum TTLs were `round numbers'. Of our entire dataset, $>$96\% were multiples of 60 seconds, and those that were not were multiples of either 15 or 20 seconds. This approach gave us high confidence that we had found the correct maximum TTL for each server domain pair.

One server was found that this approach did not work on. This server reset the cache record for an expiring domain when the record was 3-5 seconds from expiring. This was noticed early when looking for viable servers, and so the experiment was not run against this server. %

Another implementation issue we ran into was a VPN service that used multiple DNS servers, each with their own cache. When connecting to this VPN you were assigned into a private network range. Within this range was a DNS server to answer queries. The range entered was different each time you connected, regardless of the location of the VPN endpoint you choose, and the cache records of the DNS servers was not shared between different network ranges. Our method worked on these DNS servers, but ultimately we did not include them in our data as each one got few hits from other users.

\section{Findings}

Over two months we observed just under 6 million DNS queries for the 2000 domains. Some of these were accessed frequently, such as instagram.com with 18378 hits, whereas some were accessed infrequently, such as risheha.com with 1 hit. Our initial analysis of the dataset involved collating the records from the different VPNs into one large dataset, and calculating the frequency which domains were accessed over all observed VPNs. As we know the time when each TTL was due to expire and when it was refreshed (whether by us or someone else) we can calculate the total observed period per domain. We can divide this by the total observed refreshes to get an estimate of the average refresh rate per domain. 

For each domain we calculate the Poisson arrival rate, $\lambda$, of $x$, where $x$ is the number of events in a fixed time period. The Poisson distribution is chosen in our initial modelling as an appropriate function for modelling count data, under the assumption that the variance and mean of the arrival rate are equal. If we were to require more specific characterisation of features of the traffic we might prefer a more flexible distribution such as the negative binomial, which allows for \textit{overdispersion} in which the variance of the arrival rate is greater than $\lambda$. For our restricted application of ranking mean arrival rate of queries per domain, however, the Poisson is a widely-used and efficient approximation.

\vspace{+0.5em}
$P\left( x \right) = \frac{{e^{ - \lambda } \lambda ^x }}{{x!}}$
\vspace{+0.5em}

We then ranked each domain sorted by the `sample lambda' - observed mean arrival rate (distinct from the `true' arrival rate):

\vspace{+0.5em}

$\hat{\lambda} = \frac{t}{o}$

\vspace{+0.5em}

We calculated each of these to 95$\%$ confidence interval:

$b = \pm 1.96\sqrt{\hat{\lambda}/o}$

\begin{figure}
  \centering
      \includegraphics[width=0.5\textwidth]{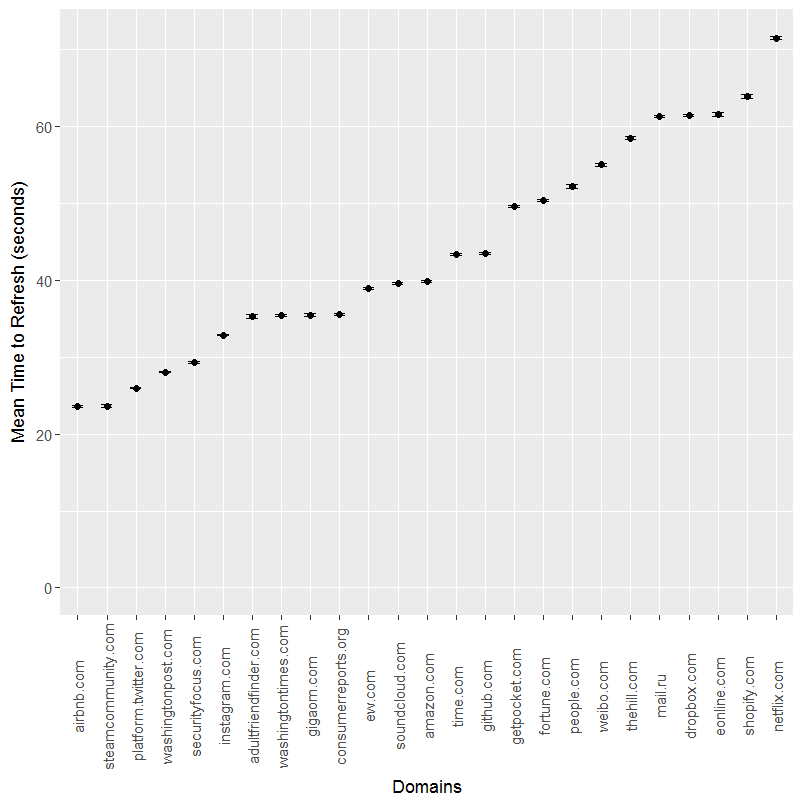}
  \caption{Most frequently accessed domains}
\end{figure}

\begin{figure}
  \centering
      \includegraphics[width=0.5\textwidth]{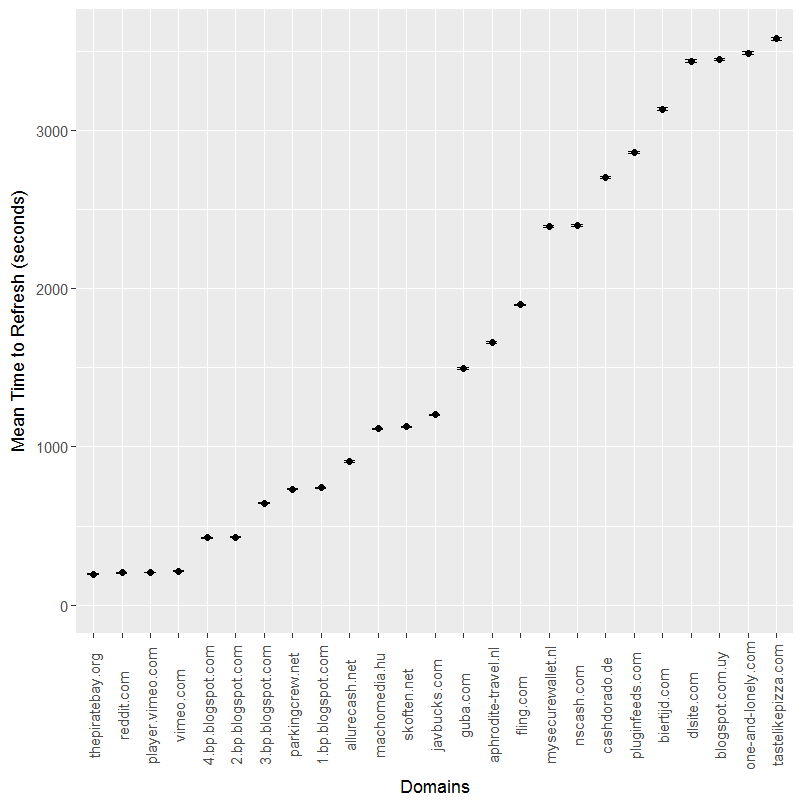}
  \caption{Most frequently accessed domains (Indonesia)}
\end{figure}

\begin{figure}
  \centering
      \includegraphics[width=0.5\textwidth]{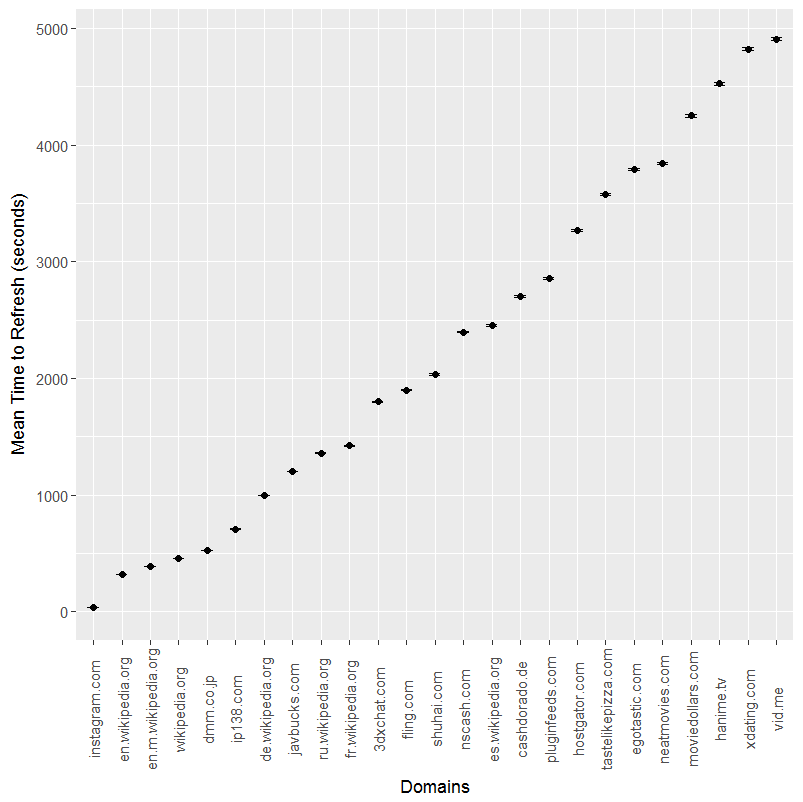}
  \caption{Most frequently accessed domains (Turkey)}
\end{figure}

\section{Discussion of Observed Domains}

Figure 1 shows the most frequently accessed domains. Popular social network sites such as twitter.com, instagram.com and weibo.com were highly represented at the top of the list. These sites encourage frequent user interaction, and also often have an app to encourage mobile use.

Some of the domains are on the list because of applications that regularly refresh their domains. High ranking examples of this are steamcommunity.com and soundhound.com, which both operate apps with a clear need for regularly communication with remote servers. Steam is a popular PC gaming application, and also offers a mobile chat application and mobile account authenticator. Soundhound is an audio recognition company, that  sends live audio samples to their remote servers in order to identify them. Other domains that fall into this category include ew.com, airbnb.com, and platform.twitter.com. These applications are often in frequent communication with their servers, which may result in a large amount of queries for these domains.

There are some unexpected domains amongst the most frequently accessed which required further investigation. securityfocus.com and gigaom.com are unlikely to be websites that people visit frequently and are not known to be associated with applications. Why are these domains being refreshed so regularly? After a further look into what these organisations do we now believe that these domains are being refreshed by automated internet research tools. Both organisations do internet research and scanning, and Security Focus also offer a range of downloadable tools. It is likely that some of these scanners or tools are 'phoning home` for information sharing. A single tool making constant requests for a particular domain has the potential to quickly rise towards the top of our most frequently refreshed domains list.

This uncertainty over what caused the query is one weakness of the approach. From just the DNS cache it is impossible to tell the cause of the query. Other causes of refreshed domains could include web crawlers, hotlinked content, or third party content or scripts embedded into other pages. Each of these would cause a request for the domain name, and from the cache alone it is impossible to know the cause. A consequence of this is that this technique only works if no one else is doing it. If this technique becomes popular amongst researchers, its results become meaningless.

However, as our approach differentiates between parent and subdomain there are some other interesting findings that shine light on the differences between automated and human requests in some circumstances. An example of this is that platform.twitter.com is accessed far more frequently than twitter.com. This subdomain is used by the Twitter API available to third parties. It seems that in our dataset the API subdomain was accessed far more frequently than the human usable site. This exemplifies a pattern amongst the data, where those domains associated with an app or tool feature more frequently than those that are only accessed manually.

We were surprised how infrequently some popular domains are accessed over the VPNs. While there are several domains which have an average refresh period of 20 seconds or less, this drops off rapidly. Of the full dataset only the 20 most frequently accessed domains had an average refresh period of less than 60 seconds. Initially we expected more domains to be refreshed as soon as they expired. Only the 103 highest frequency domains were accessed more regularly than once every 1000 seconds.

The most frequently accessed domains in China are social networking sites known to be blocked in this region. Most prominent among these are twitter.com, instagram.com, facebook.com, youtube.com, and tumblr.com. There are also several news or information sharing sites such as wsj.com, storyify.com, archive.org, wikipedia.org and blogspot.com. As our method is unable to differentiate between users attempting to access these sites from China versus the rest of the world, it is impossible for us to tell how much of this traffic is from users attempting to circumvent censorship.

When performing the language specific breakdown, only two of the countries seemed to be blocking a significant amount of local language websites. China and Iran were both found to be blocking local language sites, while Indonesia and Turkey seemed to block domains primarily written in other languages. Most of the language specific domains were accessed fairly infrequently, but there are some that stand out.
 
Of the Farsi specific domains censored in Iran in Figure 4, radiofarda.com is highly refreshed. Radio Farda is the Farsi language station of Radio Free Europe/Radio Liberty, funded by the U.S government. This was by far the most frequently accessed Iranian blocked Farsi language domain discovered, and was refreshed on average every 150 seconds with 3000 hits over the observed period. It should be noted that there is a Radio Farda mobile app, so it may be affected in a similar way to some of the domains mentioned above. Similarly, in China voachinese.com is the most frequently refreshed blocked language specific domain in Figure 5. This is operated by Voice of America, again funded by the US government. Both Radio Liberty and Voice of America are US government 'soft power`, and were recently classified as foreign agents by Russia \cite{russiavoa}. Because of their funding they are not reliant on being commercially successful in their targeted countries. It is interesting that they are still accessed frequently despite being language specific and blocked in their targeted countries. As these languages are specific to the countries where the domain is blocked in, it is more likely that users of the VPN may be doing so specifically to access blocked content.

\setcounter{totalnumber}{2}

\begin{figure}
  \centering
      \includegraphics[width=0.5\textwidth]{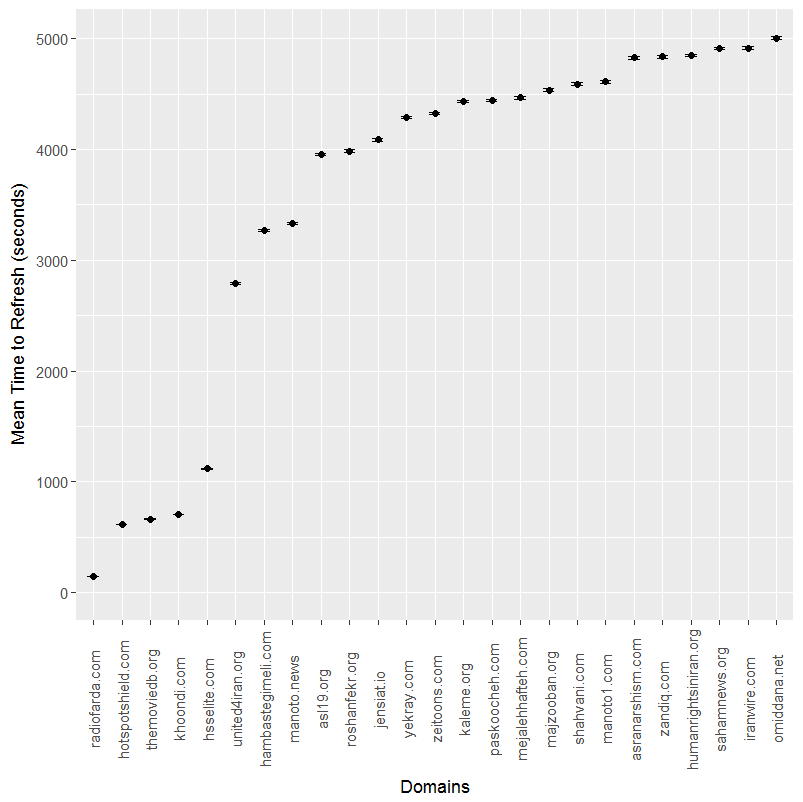}
  \caption{Most frequently accessed domains (Farsi language)}
\end{figure}

\begin{figure}
  \centering
      \includegraphics[width=0.5\textwidth]{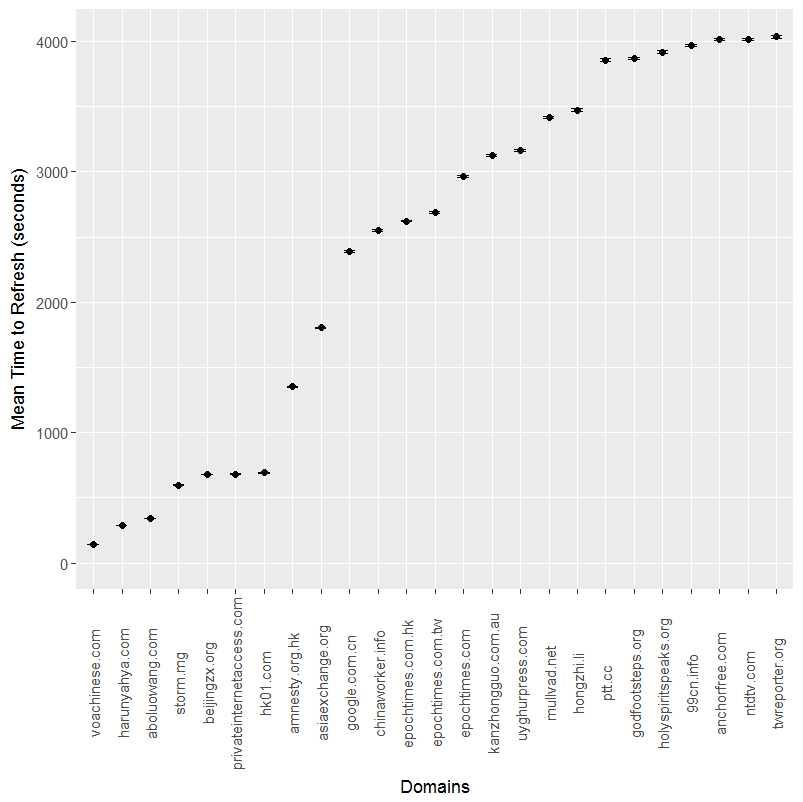}
  \caption{Most frequently accessed domains (Chinese language)}
\end{figure}

\section{Related Research}

There has not been much work on DNS cache snooping itself. A 2004 paper by Grangeia outlines the possibility of DNS cache snooping, but does not implement it or use it to measure relative popularity of domains \cite{grangeia2004dns}. In 2010 Krishnan et al. point out that a similar cache snooping technique may cause loss of privacy, specifically in the context of pre-caching web content \cite{krishnan2010dns}. The risk of DNS cache snooping is mentioned in briefly in RFC7626 \cite{rfc7626}.

Winter et al. \cite{winter2014spoiled} has worked on the privacy risk DNS queries pose to users of the Tor service. Their work looks at the risk of identifying users by the DNS requests going into and out of the Tor network. There has been some discussion over the role of VPNs as providers of privacy, anonymity and censorship resistance. One of the earliest evaluations was from Appelbaum et al. in 2012, who performed an assessment of VPNs for these purposes and found them lacking overall \cite{appelbaum2012vpwns}. In 2015 Perta et al. performed a similar evaluation, with some time spent on DNS directly \cite{perta2015glance}.

\section{Conclusions}

In this paper we present DNS cache snooping as a practical method that we use to determine how frequently domains are accessed over three large VPN services. We explored three different forms of DNS cache snooping, and discovered that cache snooping based on the TTL of DNS records is able to stably and consistently reveal information about the state of DNS records cache on a given server and how frequently these records are accessed. Using this approach, we analyse the internal DNS servers of several different VPN providers. These VPN providers advertise their service as a method of evading censorship and surveillance \cite{bypassrussia}\cite{ipvanishbypass}\cite{vpnguru}, and evidence suggests they are used for these purposes \cite{appelbaum2012vpwns}\cite{perta2015glance}. We ran this experiment for during April and May 2018, across three VPNs and 2000 domains.

During this period we discovered which of these domains were most regularly accessed through the VPN. By looking at the time it took for the records of these domains to be refreshed, we were able to calculate the overall frequency of access for each of these domains. This gave us: 1) an overall ranking of how frequently each domain was accessed through the VPN; 2) an estimate at the total number of requests for each domain through the VPN; 3) a breakdown of which sites and which type of sites are popular for users of VPNs. We did this for popular domains and again for domains we knew to be censored in specific countries: China, Indonesia, Iran, and Turkey. Two of the countries, China and Iran, were found to be blocking a large amount of websites written in their primary languages: Chinese and Farsi respectively. For these countries we provided breakdowns of the language specific blocked domains that were accessed over the VPN.

\section{Future Research}

We are interested to see if our technique can be used to discover how frequently attempts are made to access censored domains within DNS censored countries. The success of this technique depends on the specific censorship apparatus within a country. To give an example of this, the Great Firewall of China (GFW) employs DNS poisoning to prevent users from resolving DNS queries for filtered domains. Users make a query for a domain, and the GFW responds with a poisoned DNS response. Users' machines cache this result, stopping them accessing censored content.

However, the original DNS response from the authentic DNS server is still sent \cite{farnan2016poisoning}. This response can be captured to view the state of this record on the server, including the record's TTL value. By looking at this with our DNS cache snooping implementation it is possible to see how frequently users attempt to access censored domains. Preliminary results indicate this approach works, and it is possible to determine how frequently attempts are made to access censored domains within China for a chosen DNS server. This technique is not limited to China, and is likely applicable to other countries.

We have also looked at applying our work to the Tor anonymity network. Tor allows clients to make DNS requests through the network, that are then resolved by an exit node. Requests through Tor have a static TTL so it is not as simple as running it against VPN DNS servers \footnote{Another way of doing this against the Tor network may be the timing based scan against an exit node, but Tor adds even more jitter noise which would need to be removed.}. However, there is some degree of flexibility in how exit node operators handle DNS queries. Official guidance from The Tor Project suggests running caching resolvers, and if these can be accessed either from within the Tor network or the public internet this approach is feasible. This is arguably a security misconfiguration issue, although given the lack of specific guidance for DNS handling it is possible to be susceptible while following configuration guidance \cite{torrelayguide}. Indeed, we were able to proof of concept this against our own chosen-configuration exit node, but with a cursory search we were not able to find other exit nodes set up in a similar fashion. As a result we were able to generate equivalent data on Tor domain frequency usage for our exit node, but as we relied upon the configuration of our own exit node to get it we do not believe it would be ethical to publish \cite{jones2015ethical}. Greschbach et al. \cite{greschbach2016effect} wrote on the the effect of DNS on Tor's anonymity, while Phillip Winter et al. \cite{winter2014spoiled} has looked at the traffic flow identification via DNS requests sent through the Tor network.

\printbibliography

\end{document}